\def\papertitle{GRAFX: An open-source library for audio processing graphs in PyTorch}

\def\paperauthorA{Sungho Lee}
\def\paperauthorB{Marco A. Martínez-Ramírez}
\def\paperauthorC{Wei-Hsiang Liao}
\def\paperauthorD{Stefan Uhlich}
\def\paperauthorE{Giorgio Fabbro}
\def\paperauthorF{Kyogu Lee}
\def\paperauthorG{Yuki Mitsufuji}

\documentclass[twoside,a4paper]{article}

\usepackage{dafx_24}
\usepackage{amsmath,amssymb,amsfonts,amsthm}
\usepackage{etoolbox}
\usepackage{euscript}
\usepackage[T1]{fontenc}
\usepackage[utf8]{inputenc}
\usepackage{ifpdf}
\usepackage[english]{babel}
\usepackage{stfloats}

\usepackage{subfig} 
\usepackage{caption}
\usepackage{xcolor}
\usepackage{booktabs}
\usepackage{colortbl}
\usepackage{multirow}
\usepackage{enumitem}
\usepackage{siunitx}
\usepackage{musicography}
\usepackage{fancyvrb}
\usepackage{listings}
\definecolor{codegreen}{rgb}{0,0.6,0}
\definecolor{codegray}{rgb}{0.5,0.5,0.5}
\definecolor{codepurple}{rgb}{0.58,0,0.82}
\definecolor{backcolour}{rgb}{0.96,0.96,0.92}

\makeatletter
\lst@AddToHook{OnEmptyLine}{\addtocounter{lstnumber}{-1}}
\makeatother

\lstdefinestyle{mystyle}{
    backgroundcolor=\color{backcolour},   
    commentstyle=\color{codegreen},
    keywordstyle=\color{magenta},
    numberstyle=\tiny\color{codegray},
    stringstyle=\color{codepurple},
    basicstyle=\ttfamily\footnotesize,
    breakatwhitespace=false,         
    breaklines=true,                 
    captionpos=b,                    
    keepspaces=true,                 
    numbers=left,   
    numberstyle=\ttfamily\scriptsize,
    numbersep=5pt,                  
    showspaces=false,                
    numberblanklines=false,
    showstringspaces=false,
    showtabs=false,                  
    tabsize=2,
        literate={0}{{\textcolor{blue}{0}}}{1}%
             {1}{{\textcolor{blue}{1}}}{1}%
             {2}{{\textcolor{blue}{2}}}{1}%
             {3}{{\textcolor{blue}{3}}}{1}%
             {4}{{\textcolor{blue}{4}}}{1}%
             {5}{{\textcolor{blue}{5}}}{1}%
             {6}{{\textcolor{blue}{6}}}{1}%
             {7}{{\textcolor{blue}{7}}}{1}%
             {8}{{\textcolor{blue}{8}}}{1}%
             {9}{{\textcolor{blue}{9}}}{1}%
}

\usepackage{algorithm}
\usepackage{algpseudocode}
\algrenewcommand\algorithmicrequire{\textbf{Input:}}
\algrenewcommand\algorithmicensure{\textbf{Output:}}
\algnewcommand\algorithmicinput{\textbf{Input:}}
\algnewcommand\algorithmicoutput{\textbf{Output:}}
\algnewcommand\Input{\item[\algorithmicinput]}%
\algnewcommand\Output{\item[\algorithmicoutput]}%

\renewcommand\paragraph[1]{\noindent\textbf{#1 ---}}

\newcommand\bfT{\mathbf{T}}
\newcommand\bbfT{\bar{\mathbf{T}}}

\newcommand\bfIG{\mathbf{I}^\mathrm{G}}
\newcommand\bfIA{\mathbf{I}^\mathrm{A}}
\newcommand\bfIP{\mathbf{I}^\mathrm{P}}
\newcommand\bfIS{\mathbf{I}^\mathrm{S}}
\newcommand\bfE{\mathbf{E}}
\newcommand\bfU{\mathbf{U}}
\newcommand\bbfU{\bar{\mathbf{U}}}
\newcommand\bfS{\mathbf{S}}
\newcommand\bfP{\mathbf{P}}
\newcommand\bfY{\mathbf{Y}}

\newcommand\fn[1]{\mathrm{#1}\:\!}
\newcommand\textttt[1]{\text{\scriptsize$\texttt{#1}$}}

\input glyphtounicode
\ninept

\newcounter{numauth}
\newcounter{listcnt}
\newcommand\authcnt[1]{\ifdefined#1 \stepcounter{numauth} \fi}

\newcommand\addauth[1]{
\ifdefined#1 
\stepcounter{listcnt}
\ifnum \value{listcnt}<\value{numauth}
\appto\authorslist{, #1}
\else
\appto\authorslist{~and~#1}
\fi
\fi}
\authcnt{\paperauthorB}
\authcnt{\paperauthorC}
\authcnt{\paperauthorD}
\authcnt{\paperauthorE}
\authcnt{\paperauthorF}
\authcnt{\paperauthorG}
\authcnt{\paperauthorH}
\authcnt{\paperauthorI}
\authcnt{\paperauthorJ}
\def\authorslist{\paperauthorA}
\addauth{\paperauthorB}
\addauth{\paperauthorC}
\addauth{\paperauthorD}
\addauth{\paperauthorE}
\addauth{\paperauthorF}
\addauth{\paperauthorG}
\addauth{\paperauthorH}
\addauth{\paperauthorI}
\addauth{\paperauthorJ}

\usepackage{times}

\newif\ifpdf
\ifx\pdfoutput\relax
\else
\ifcase\pdfoutput
\pdffalse
\else
\pdftrue
\fi

\ifpdf 
\usepackage[pdftex,
pdftitle={\papertitle},
pdfauthor={\authorslist},
pdfsubject={Proceedings of the 27th International Conference on Digital Audio Effects (DAFx24)},
colorlinks=false, 
bookmarksnumbered, 
pdfstartview=XYZ 
]{hyperref}
\usepackage[pdftex]{graphicx}
\else 
\usepackage[dvips]{epsfig,graphicx}
\usepackage[dvips,
pdftitle={\papertitle},
pdfauthor={\authorslist},
pdfsubject={Proceedings of the 27th International Conference on Digital Audio Effects (DAFx24)},
colorlinks=false, 
bookmarksnumbered, 
pdfstartview=XYZ 
]{hyperref}
\fi
\usepackage[hypcap=true]{caption}
\usepackage[edges]{forest}

\title{\papertitle}

\affiliation{
\hspace{-4.5mm}
\paperauthorA$^\dagger$\sthanks{Work partially done during an internship at Sony AI. 
},\, 
\paperauthorB$^\natural$,\, \paperauthorC$^\natural$,\, \paperauthorD$^\sharp$,\, \paperauthorE$^\sharp$,\, \paperauthorF$^\dagger$,\, and \paperauthorG$^{\natural\flat}$
}
{
\hspace{-5mm}
$^\dagger$Department of Intelligence and Information, Seoul National University, Seoul, South Korea \\
\hspace{-5mm}
$^\natural$Sony AI, Tokyo, Japan \quad
$^\sharp$Sony Europe B.V., Stuttgart, Germany \quad 
$^\flat$Sony Group Corporation, Tokyo, Japan
}
\definecolor{emerald}{rgb}{0.31, 0.78, 0.47}

\usepackage{tikz}
\usetikzlibrary{external}
\usetikzlibrary{arrows.meta, bending, decorations.pathmorphing}

\begin{document}
\ifpdf 
\DeclareGraphicsExtensions{.png,.jpg,.pdf}
\else  
\DeclareGraphicsExtensions{.eps}
\fi

\makeatletter
\pdfbookmark[0]{\@pdftitle}{title}
\makeatother

\maketitle
\begin{abstract}
We present \texttt{GRAFX}, an open-source library designed for handling audio processing graphs in \texttt{PyTorch}. 
Along with various library functionalities, 
we describe technical details on the efficient parallel computation of input graphs, signals, and processor parameters in GPU.
Then, we show its example use under a music mixing scenario, where parameters of every differentiable processor in a large graph are optimized via gradient descent.
The code is available at \url{https://github.com/sh-lee97/grafx}.
\end{abstract}

\vspace{-1mm}
\section{Introduction}
\vspace{-1mm}
\paragraph{Motivation}
\looseness=-1
Under the umbrella of so-called \emph{differentiable signal processing} \cite{engel2020ddsp, hayes2023review},
numerous attempts have been made to import existing audio processors to automatic differentiation frameworks, e.g., \texttt{PyTorch} \cite{paszke2019pytorch}.
Differentiable processors, as standalone modules, allow gradient-based optimization of their parameters. 
Alternatively, they can be used to train a neural network as a parameter estimator \cite{lee2022differentiable, steinmetz2022style}.
In either case, as the processors are identical to or approximate the real-world ones, the obtained parameters are easy to interpret and control. 
To further leverage this advantage of differentiable signal processing, it would be desirable to consider the \emph{composition} of processors since it is a standard real-world practice. 
In a more general setting, this composition can be represented in a \emph{graph} format \cite{lee2023blind}.
Yet, there are few public implementations \cite{engel2020ddsp, uzrad2024diffmoog} that provide highly flexible graph-related functionalities.

\vspace{1.5mm}
\paragraph{Contributions}
\looseness=-1
In response, we present a library called \texttt{GRAFX} that allows users to handle audio processing graphs and their applications in \texttt{PyTorch}.
Along with the open-source code, this demonstration paper serves multiple purposes. First, we highlight the library's core components and applications.
This includes our custom data structures that allow the creation and modification of graphs, computation of output signals, and potential use as input of graph neural networks (GNNs) \cite{lee2023blind}.
Second, we dive into the heart of \texttt{GRAFX}, an optimized processing algorithm that computes the output audio from graphs, source audio, and processor parameters. 
Unlike the previous implementations \cite{engel2020ddsp, uzrad2024diffmoog},
ours allows the change of the graph for every optimization step. This is useful in multiple scenarios, e.g., training a GNN that predicts the parameters of given graphs \cite{lee2023blind} or pruning a graph with gradient descent \cite{lee2024searching}.
Also, we use batched node processing faster than the conventional ``one-by-one'' computation of processors. 
Third, our library is complemented with various differentiable processors; we report their technical details. Finally, we describe how we used \texttt{GRAFX} to create the music mixing graphs in our companion paper \cite{lee2024searching}
and evaluate the speedup we obtain with the batched node processing.

\section{Audio processing graphs in GPU}\label{section:diff_apg}
\vspace{-1mm}
\paragraph{Definitions}
We write an audio processing graph as $G=(V, E)$ where $V$ and $E$ are the node and edge set, respectively. 
Each node $v_i \in V$ can represent a processor $f_i$ with a type $t_i$, e.g., reverb \texttt{r}. It takes $M$ signal(s) $u_i[n]$ and a collection of parameter tensors $p_i$ as input and produces $N$ output(s) $y_i[n]$ ($n$ denotes a time index).
\begin{subequations}
\begin{align}
    {y_{i}^{(1)}[n], \cdots, y_{i}^{(N)}[n]}
    &=  f_i\big(
    {u_{i}^{(1)}[n], \cdots, u_{i}^{(M)}[n]}
    , p_i\big). \\
    u_{i}^{(l)}[n] &=  \sum_{(j, k) \in \mathcal{N}^+(i, l)} y_{j}^{(k)}[n].
\end{align}
\end{subequations}
\looseness=-1
Here, $\mathcal{N}^+(i, l)$ denotes a collection of nodes and channel indices that send their output signals to $l^\mathrm{th}$ input of node $i$. 
With this setup, each edge $e_{ij}\in E$ becomes a ``cable'' connecting two nodes. Note that each edge also requires a type attribute $t_{ij}=(k, l)$, a tuple of input and output channel indices, unless every processor in a graph is a single-input single-output system (SISO), i.e., $M=N=1$.
In short, each node’s outputs can be computed by finding its inputs, aggregating those, and processing the sums with the parameters. 
The graph output can be obtained by repeating this
procedure over all nodes in topological order, starting from input nodes \texttt{i} until we reach the output nodes \texttt{o}. 
We allow any graph except those with cycles, as feedback loops are hard to resolve and bottleneck the processing speed due to the forced
sample-level recursion.

\vspace{1.5mm}
\paragraph{Graph representations}
\looseness=-1
There are two main use scenarios when handling the audio processing graphs.
First, users may create and modify a graph. For this case, we provide a mutable data structure \texttt{GRAFX} (same as the library name). It inherits \texttt{MultiDiGraph} class from \texttt{networkx} \cite{hagberg2008exploring} and provides additional functionalities, e.g., adding a serial chain of processors.
Second, each graph can be used to compute output audio or fed into a GNN. In such cases, representing each graph as a collection of tensors is more convenient and efficient. Therefore, we provide a \texttt{GRAFXTensor} class, which is compatible with \texttt{Data} class from \texttt{torch\_geometric} \cite{fey2019fast}.
The following are the tensors we use to describe each graph.
First, we have a node type vector $\bfT_V \in \mathbb{N}^{|V|}$, an edge index tensor $\bfE \in \mathbb{N}^{2\times|E|}$, and an (optional) edge type tensor $\bfT_E \in \mathbb{N}^{2\times |E|}$ where $|\cdot |$ denotes the size of a given set.
All parameters are collected in a dictionary $\bfP$
whose key is a node type $t$ and value $\bfP[t]$ contains the parameters of that type.
In this paper, we assume that the value is a single tensor in a form $\bfP[t]\in \mathbb{R}^{|V_t|\times N_t}$ 
where $|V_t|$ and $N_t$ are the number of nodes and parameters.
In fact, we allow $\bfP[t]$ to be a tensor with more dimensions or even a dictionary of tensors; we omit such cases for simplicity.
All sources are stacked to a single tensor $\mathbf{S} \in \mathbb{R}^{K\times C\times L}$ where $K$, $C$, and $L$ are the number of sources, channels, and length, respectively. 
We ensure that all the tensors, 
$\bfT_V$, $\bfT_E$, $\bfE$, $\bfP$, and $\bfS$, 
share the same node order.
For example, 
a $k^\mathrm{th}$ source $s_k[n]$ must correspond to the first $k^\mathrm{th}$ input \texttt{i} 
in the node type list $\bfT_V$.
Likewise, an $l^\mathrm{th}$ type-$t$ parameter $\bfP[t]_l \in \mathbb{R}^{N_t}$ 
must correspond to the $l^\mathrm{th}$ type $t$ in the type list $\bfT_V$.

\begin{figure}
    \vspace{-3.5mm}
    \captionsetup[subfloat]{captionskip=4pt}
    \centering
    \subfloat[\it Target graph \label{fig:full}]{
        \includegraphics[width=.49\columnwidth]{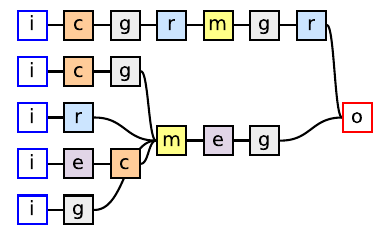}} 
    \subfloat[\it Optimal schedule \label{fig:oracle}]{
        \includegraphics[width=.49\columnwidth]{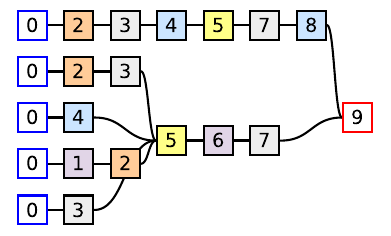}}\\
    \vspace{-1mm}
    \subfloat[\it Greedy schedule \label{fig:greedy}]{
        \includegraphics[width=.49\columnwidth]{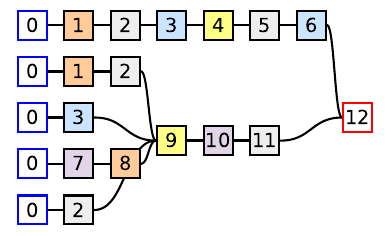}}
    \subfloat[\it One-by-one processing \label{fig:1by1}]{
        \includegraphics[width=.49\columnwidth]{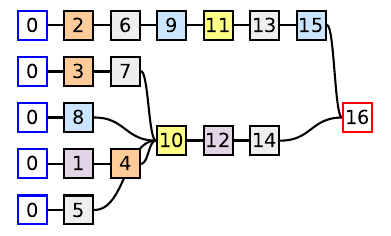}} \\
    \vspace{1mm}
    {\footnotesize \it \texttt{i}: input, \texttt{o}: output, \texttt{m}: mix, \texttt{e}: equalizer, \texttt{c}: compressor, \texttt{g}: gain/panning.
    }
    \vspace{-4mm}
\caption{
\it Various schedules for batched node processing. 
For each schedule, the processing orders are shown inside the nodes. 
}
  \label{fig:schedule} 
\vspace{-4mm}
\end{figure}

\paragraph{Batched processing}
\looseness=-1
For faster computation in GPU, 
maximizing the parallelism is desirable, and batched processing is the most standard approach. Note that we have $3$ levels of batched processing. 
First, we may want \emph{source-level} parallelism, i.e., processing batches of multiple input sources with a single graph. 
This can be easily achieved when every processor supports batched processing.
Next, we can consider \emph{node-level} parallelism, processing multiple nodes with the same type simultaneously.
Specifically, consider a sequence of $N+1$ node subsets
$V_0, \cdots, V_N \subset V$ satisfying the following conditions.
\begin{enumerate}[label=(\roman*), leftmargin=7mm]
  \setlength\itemsep{0em}
    \item 
		It forms a \emph{partition}: $\cup_n V_n = V$ and $V_n\cap V_m = \emptyset$ if $n \neq m$.
    \item 
		It is \emph{causal}:
		no path from $u \in V_n$ to $v \in V_m$ exists
		if $n \geq m$.
    \item 
		Each subset $V_n$ is \emph{homogeneous}: it has only a single type $t_n$.
\end{enumerate}
Then, we can compute a batch of output signals $\bfY_n$ of each subset $V_n$ sequentially, from $n=0$ to $N$.
Consequently, we reduce the number of the gather-aggregate-process iterations from $|V|$ to $N$ (we have no processings for $n=0$ as $V_0$ contains input modules). 
Figure \ref{fig:schedule} shows an example.
For a graph with $|V|=21$ nodes (\ref{fig:full}), 
we can obtain a sequence with $N=9$ (\ref{fig:oracle}). 
Finally, equipped with this batched node processing, we can also achieve \emph{graph-level} parallelism; we can simply treat a batch of multiple graphs as a single large disconnected graph. 
Therefore, we will focus on the batched \emph{node} processing for the remainder of this section.

\begin{figure}
\centering
\captionsetup[subfloat]{captionskip=0pt}
\subfloat[\it An example code for creating and rendering a mixing console.  \label{fig:code}]{
\resizebox{0.96\columnwidth}{!}{
\parbox{1.05\columnwidth}{
\begin{lstlisting}[language=python, escapechar=|]
import torch |\label{code:import-torch}|
from grafx import * |\label{code:import-grafx}|                                 

G = GRAFX() |\label{code:declare}|
in_id = G.add("in") |\label{code:add-in}|
chain = ["eq", ..., "reverb"] |\label{code:chain}|
start_id, end_id = G.add_serial_chain(chain) |\label{code:add-chain}|
mix_id = G.add("mix") |\label{code:add-mix}|
G.connect(in_id, start_id) |\label{code:connect-in}|
G.connect(end_id, mix_id) ... |\label{code:connect-mix}|

G_t = convert_to_tensor(G) |\label{code:tensor}|
render_data = compute_render_data(G_t) |\label{code:render}|
source = torch.rand(1, 5, 2, 2**17) |\label{code:source}|
processors = {"eq": ZerophaseFIREqualizer(), ...   "reverb": MideSideFilteredNoiseReverb()} |\label{code:processors}|
parameters = {"eq": torch.zeros(7, 1024), ...      "reverb": torch.zeros(7, 768)} |\label{code:parameters}|
output = render_grafx(source, processors, parameters, render_data) |\label{code:render}|
\end{lstlisting}
}
}} \\
\vspace{-1.5mm}
\subfloat[\it \centering An example mixing console, drawn with \texttt{draw\_grafx}. 
 The same notation as Fig. \ref{fig:full} plus \texttt{n}: noisegate, \texttt{s}: stereo imager, \texttt{d}: multitap delay. 
\label{fig:example-console}]{
\parbox{.98\columnwidth}{
\hspace{4mm}
\includegraphics[height=2.15cm]{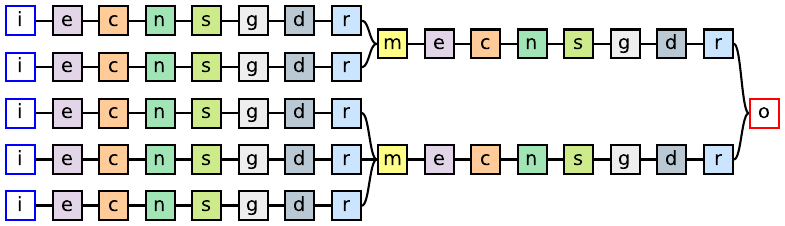}
\vspace{1.5mm}
}
}
\vspace{-5mm}
\caption{\it 
Example usage of \texttt{GRAFX} for a music mixing scenario. 
}
\label{fig:grafx-example}
\vspace{-3.4mm}
\end{figure}

\vspace{1.5mm}
\paragraph{Type scheduling}
\looseness=-1
For a maximized node-level parallelism, we want to find the shortest node subset sequence.
This is a variant of the scheduling problem. 
First, we always choose a maximal subset $V_i$ when the type $t_i$ is fixed. 
This makes the subset sequence equivalent to a type string, 
e.g., \texttt{iecgrmegro} for \ref{fig:oracle}.
We also choose the first and the last subset, $V_0$ and $V_N$, 
to have all of the input and output nodes, respectively.
Since the search tree for the shortest sequence exponentially grows,
the brute-force search is too expensive for most graphs. 
Instead, we may try the {greedy} method
that chooses a type with the largest number of computable nodes (\ref{fig:greedy}).
However, this usually results in a longer sequence, thus slower processing.
We can alleviate this issue with the beam search, i.e., keeping multiple best schedules as candidates instead of one.  
Intuitively, the batched node processing is effective for graphs with fewer types and a certain structure, e.g., ones that apply processors in the same order for every input (e.g., see Figure \ref{fig:example-console}).

\begin{figure}
\vspace{-3.2mm}
\begin{algorithm}[H]
\caption{\it Batch computation of audio processing graphs.}\label{alg:cap}
\begin{algorithmic}[1]
\Require Types $\bfT_V$ and $\bfT_E$, edges $\bfE$, parameters $\bfP$, and inputs $\bfS$
\Ensure Output signals $\bfY$ and (optional) intermediate signals $\bfU$
\vspace{.7mm}
\State $\bbfT, N \gets \fn{ScheduleBatchedProcessing}(\bfT_V, \bfE)$ \label{algo:render:schedule}
\State $\sigma \gets \fn{OptimizeNodeOrder}(\bbfT, \bfT_V, \bfE)$ \label{algo:render:find-order}
\State $\bfT_V, \bfE, \bfP \gets \fn{Reorder}(\sigma, \bfT_V, \bfE, \bfP)$ \label{algo:render:reorder}
\State $\bfIG, \bfIP, \bfIA, \bfIS \gets \fn{GetReadWriteIndex}(\bbfT, \bfT_V, \bfE, \bfT_E, \bfP)$ \label{algo:render:calculate_index}
\State $\bfU \gets \fn{Initialize}(\bfS, \bfT_V)$ \label{algo:render:initialize}
\For{$n$ $\gets$ $1$ to $N$} \label{algo:render:iter-start}
	\State $\bbfU_n \gets \fn{Gather}(\bfU, \bfIG_n)$ 
	\Comment{\texttt{index\_select}}
	\label{algo:render:gather-input}
    \State $\bfU_n \gets \fn{Aggregate}(\bbfU_n, \bfIA_n)$ 
	\Comment{\texttt{scatter}}
	\label{algo:render:aggregate}
	\State $\bfP_n \gets \fn{Gather}(\bfP[\bar{t}_n], \bfIP_n)$ 
	\Comment{\texttt{slice}}
	\label{algo:render:gather-param}
	\State $\mathbf{Y}_n \gets \fn{Process}(\bar{t}_n, \bfU_n, \bfP_n)$ 
	\label{algo:render:process}
	\State $\bfU \gets \fn{Store}(\mathbf{U}, \mathbf{Y}_n, \bfIS_n)$ 
	\Comment{\texttt{slice}}
	\label{algo:render:store}
\EndFor \label{algo:render:iter-end}
\State $\bfY \gets \bfY_N$ \label{algo:render:out}
\State \Return $\bfY, \bfU$
\end{algorithmic}
\label{algo:rendering} 
\end{algorithm}
\vspace{-7.3mm}
\end{figure}

\paragraph{Implementation details}
\looseness=-1
Algorithm \ref{algo:rendering} obtains the output 
$\bfY$ 
from the prescribed inputs (inside the following parentheses denote the line numbers).
First, we schedule the batched node processing and obtain a node type list $\bbfT \in \mathbb{N}^{N+1}$ 
(\ref{algo:render:schedule}).
Next, as the main batched processing loop 
(\ref{algo:render:iter-start}-\ref{algo:render:iter-end}) contains multiple memory reads/writes, 
we calculate the node reordering $\sigma$ 
that achieves contiguous memory accesses and 
improves the computation speed
(\ref{algo:render:find-order}). 
This procedure allows memory accesses via \texttt{slice}, 
as shown in the comments of Algorithm \ref{algo:rendering}.
After reordering the graph tensors with $\sigma$ (\ref{algo:render:reorder}),
we retrieve lists of indices, $\bfIG$, $\bfIP$, $\bfIA$, and $\bfIS$, 
used for the tensor read/writes in the main loop (\ref{algo:render:calculate_index}).
Note that all these steps 
(\ref{algo:render:schedule}-\ref{algo:render:calculate_index})
are done in CPU and, in most cases, in multiple separate threads.
Therefore, they do not bottleneck the GPU and optimization.
After the preprocessing, we create an intermediate output tensor 
$\bfU$, which will have a shape of $|V|\times C\times L$ if all processors are SISO systems or $N_\mathrm{sum} \times C\times L$ where $N_\mathrm{sum}$ denotes the total number of outputs in the graph (\ref{algo:render:initialize}).
As we put all the inputs to be the first partition $V_0$, 
it can be initialized with simple concatenation: $\bfU = \bfS \oplus \mathbf{0}$.
The remaining repeats batched processing and necessary reads/writes 
(\ref{algo:render:iter-start}-\ref{algo:render:iter-end}).
For each $n^\mathrm{th}$ iteration, 
we collect the previous outputs $\bbfU_n$ that are routed to the current partition nodes. 
We achieve this by accessing the intermediate tensor $\bfU$ with the index $\bfIG_n$
with \texttt{index\_select}
(\ref{algo:render:gather-input}).
Then, we aggregate them using \texttt{scatter}
if multiple edges are connected to some nodes
(\ref{algo:render:aggregate}).
We can similarly obtain a parameter tensor $\bfP_n$
with its corresponding index $\bfIP_n$. 
Especially, our node reordering (\ref{algo:render:reorder}) 
makes this a simple \texttt{slice}, faster than the usual \texttt{index\_select}.
With the obtained input signals $\bfU_n \in \mathbb{R}^{|V_n|\times C\times L}$ and parameters $\bfP_n \in \mathbb{R}^{|V_n|\times N_t}$, 
we batch-compute the node outputs $\bfY_n$ (\ref{algo:render:process}).
Then, we save them to the intermediate output tensor $\bfU$ with the \texttt{slice} index $\bfIS_n$ (\ref{algo:render:store}) so that the remaining steps can access them as inputs.
After the iteration, we have all node outputs saved in $\bfU$. 
The final graph outputs are given as output of the last step $\bfY = \bfY_N \in \mathbb{R}^{|V_N|\times C\times L}$ 
since we set the last node partition $V_N$ to collect all output nodes.

\vspace{-1mm}
\section{Differentiable Audio Processors} \label{section:processor-details}
\vspace{-1mm}
We report the differentiable audio processors that we provide. 
By default, they accept and produce stereo signals, i.e., $C=2$.
All the hyperparameters, e.g., the number of filter taps, are ones used in the companion paper \cite{lee2024searching}.
We use \texttt{FlashFFTConv} \cite{fu2023flashfftconv} for every causal convolution to speed up the processing and save memory.

\paragraph{Gain/panning}
We use simple channel-wise constant multiplication. 
Its parameter vector $p_\textttt{g} \in \mathbb{R}^2$ is in log scale, 
so we apply exponentiation before multiplying it to the stereo signal. 
\begin{equation}
    y[n] = \exp (p_\textttt{g}) \cdot u[n].     
\end{equation}

\vspace{1mm}
\paragraph{Stereo imager}
We multiply the side signal, i.e., left minus right, with a gain parameter $p_\textttt{s} \in \mathbb{R}$ 
to control the stereo width. The mid and side outputs are given as
\begin{subequations}
\begin{align}
    y_\mathrm{m}[n] &= u_\mathrm{l}[n] + u_\mathrm{r}[n], \\
    y_\mathrm{s}[n] &= \exp (p_\textttt{s}) \cdot (u_\mathrm{l}[n] - u_\mathrm{r}[n]).
\end{align}
\end{subequations}
Then, we convert the mid/side output to a stereo signal back as follows, 
$y_\mathrm{l}[n] = (y_\mathrm{m}[n]+y_\mathrm{s}[n])/2$ and $y_\mathrm{r}[n] = (y_\mathrm{m}[n]-y_\mathrm{s}[n])/2$.

\vspace{2mm}
\paragraph{Equalizer}
We use a single-channel zero-phase FIR filter. 
Considering its log-magnitude as a parameter $p_\textttt{e}$, 
we compute inverse FFT (IFFT) of the magnitude response and multiply it with a Hann window $v^\mathrm{Hann}[n]$.
As a result, the length-$N$ FIR is given as 
\begin{equation}
    h_\textttt{e}[n] = v^\mathrm{Hann}[n] \cdot \frac{1}{N} \sum_{k=0}^{N-1} \exp p_\textttt{e}[k] \cdot  w_{N}^{kn}
\end{equation}
where $-(N+1)/2 \leq n \leq (N+1)/2$ and $w_{N} = \exp(j\cdot 2\pi/N)$.
We compute the final output by applying the same FIR to both the left and right channels as follows, 
\begin{equation}
    y_\mathrm{x}[n] = u_\mathrm{x}[n]*h_\textttt{e}[n] \quad (\mathrm{x} \in \{\mathrm{l}, \mathrm{r}\}).
\end{equation}
We set the FIR length to $N=2047$. Therefore, the parameter $p_\textttt{e}$ has a size of $1024$.

\vspace{2mm}
\paragraph{Reverb}
We use a variant of the filtered noise model \cite{engel2020ddsp}. 
First, we create $2$ seconds of uniform noises, $u_\mathrm{m}[n]$ and $u_\mathrm{s}[n]$.
Next, we apply a magnitude mask $M_\mathrm{x}[k, m]$ to each noise's STFT $U_\mathrm{x}[k, m]$ as follows, 
\begin{equation}
    H_\mathrm{x}[k, m] = U_\mathrm{x}[k, m] \odot M_\mathrm{x}[k, m] \quad (\mathrm{x} \in \{\mathrm{m}, \mathrm{s}\}).
\end{equation}
where $k$ and $m$ denote frequency and time frame index.
Each mask is parameterized with an initial coloration $H^0_\mathrm{x}[k]$ and an absorption filter $H^\Delta_\mathrm{x}[k]$ both in log magnitudes as follows,
\begin{equation}
	M_\mathrm{x}[k, m] = \exp ({H^0_\mathrm{x}[k] + (m-1) H^\Delta_\mathrm{x}[k]}).
\end{equation}
Next, we convert the masked STFTs to the time-domain responses, $h_\mathrm{m}[n]$ and $h_\mathrm{s}[n]$.
We obtain the desired FIR $h_\textttt{r}[n]$ by converting the mid/side to stereo.
We apply channel-wise convolutions to the input $u[n]$ and get the output $y[n]$.
The FFT and hop lengths are $384$ and $192$, respectively.
The parameter $p_\textttt{r}$ has a size of $768$ ($2$ channels, each with $2$ filters with $192$ magnitudes).

\vspace{2mm}
\paragraph{Compressor}
We implement the canonical feed-forward digital compressor
\cite{giannoulis2012digital}.
First, for a given input audio, we sum the left and right channels to obtain a mid signal $u_\mathrm{m}[n]$. Then, we calculate its energy envelope $G_u[n] = \log g_u[n]$ where
\begin{equation}\label{eq:envelope}
	g_u[n] = \alpha[n] g_u[n-1]+(1-\alpha[n]) u_\mathrm{m}^2[n].
\end{equation}
Here, the coefficient $\alpha[n]$ is typically set to a different constant for an ``attack'' (where $g_u[n]$ increases) and ``release'' (where $g_u[n]$ decreases) phase.
As this part (also known as ballistics) bottlenecks the computation speed in GPU,
following the recent work \cite{steinmetz2022style}, we restrict the coefficients to the same value $\alpha$.
By doing so, Equation \ref{eq:envelope} simplifies to a one-pole IIR filter, whose impulse response up to a finite length $N$ can be exactly obtained in parallel as follows,
\begin{equation} \label{eq:ballistics-approximation}
	h^\mathrm{env}[n] = (1-\alpha) \alpha^n.
\end{equation}
Therefore, the energy envelope $g_u[n]$ can be simply computed as a convolution between the FIR $h^\mathrm{env}[n]$ and the energy signal $\smash{u^2_\mathrm{m}[n]}$.
Next, we calculate the compressed energy envelope $\smash{G_y[n]}$.
We use a quadratic knee, interpolating the compression and the bypass region.
For a given threshold $T$ and  half of the knee width $W$, 
\begin{equation}
	G_y[n] = \begin{cases}
		G_y^\mathrm{above}[n] & G_u[n] \geq T+W,  \\
		G_y^\mathrm{mid}[n]   & T-W \leq G_u[n] < T+W, \\
		G_y^\mathrm{below}[n] & G_u[n] < T-W
		\end{cases}
\end{equation}
where, for a given compression ratio $R$, each term is 
\begin{subequations}
\begin{align}
	G_y^\mathrm{above}[n] &= T+\frac{G_u[n]-T}{R}, \\
	G_y^\mathrm{mid}[n]   &= G_u[n] + \Big(\frac{1}{R}-1\Big)\frac{(G_u[n]-T+W)^2}{4W},
\end{align}
\end{subequations}
and $G_y^\mathrm{below}[n] = G_u[n]$. 
Finally, we can compute the output as
\begin{equation}
	y_\mathrm{x}[n] = \exp(G_y[n]-G_u[n]) \cdot u_\mathrm{x}[n] \quad (\mathrm{x} \in \{\mathrm{l}, \mathrm{r}\}).
\end{equation}
The scalar parameters introduced above, $\alpha$, $T$, $W$, and $R$, are concatenated and used as a parameter vector  $p_\textttt{c} \in \mathbb{R}^4$.

\vspace{2mm}
\paragraph{Noisegate}
It is identical to the compressor above, except for the gain computation: we set
$G_y^\mathrm{above}[n] = G_u[n]$ and 
\begin{subequations}
\begin{align}
	G_y^\mathrm{mid}[n]   &= G_u[n] + (1-R)\frac{(G_u[n]-T-W)^2}{4W}, \\
	G_y^\mathrm{below}[n] &= T+R(G_u[n]-T).
\end{align}
\end{subequations}

\vspace{2mm}
\paragraph{Multitap delay}
We use a $2$ seconds of delay effect with at most one delay $d_m$ at every $100\si{ms}$
(therefore, the number of delay taps is $M=20$). 
Each delay is filtered with an FIR $c_m[n]$, parameterized in the same way as the zero-phase equalizer but with $39$ taps. 
Separate delays and filters are used for each left and right channel, but we will omit this for simplicity.
Under this setting, the multitap delay's FIR is given as follows,
\begin{equation}
	h_\textttt{d}[n] = \sum_{m=1}^{M} c_m[n]*\delta[n-d_m]
\end{equation}
where $\delta[n]$ is an unit impulse.
Here, we aim to optimize each delay length $d_m \in \mathbb{N}$, which is discrete, using gradient descent.
To this end, we exploit the fact that each delay $\delta[n-d_m]$ corresponds to a complex sinusoid in the frequency domain.
Recent work showed that the sinusoid's complex angular frequency $z_m \in \mathbb{C}$ can be optimized with the gradient descent 
when we allow it to be inside the unit disk, i.e., $|z_m| \leq 1$ \cite{hayes2023sinusoidal}. 
We leverage this finding; for each delay, we compute a damped sinusoid with the angular frequency $z_m$. Then, we use its inverse FFT as a surrogate of the delay signal.
\begin{equation}
	\delta[n-d_m] \approx \frac{1}{N} \sum_{k=0}^{N-1} z_m^k w_N^{kn}.
\end{equation}
As this signal only approximates the exact delay,
we use it only for backpropagation with the straight-through estimation \cite{bengio2013estimating}. 
Moreover, we normalize each gradient and regularize
the parameter $z_m$ to be closer to the unit circle.
This processor has a parameter $p_\textttt{d}$ of size $880$ ($40$ delays, each delay with a complex angular frequency and $20$ log-magnitudes of an equalizer). 

\section{Music Mixing Applications}
\vspace{-1mm}
\paragraph{Example usage}
\looseness=-1
Figure \ref{fig:code} demonstrates how we used \texttt{GRAFX} to construct a graph called ``music mixing console'' in our companion paper \cite{lee2024searching}. Note that this is a modified and simplified version of the original code (inside the following parentheses denote the line numbers).
First, we import the libraries (\ref{code:import-torch}-\ref{code:import-grafx}). Then, we create an empty graph (\ref{code:declare}) and add the necessary nodes. 
We add a single input node \texttt{in} with \texttt{add} (\ref{code:add-in}). We also add a serial chain of processors by passing a sequence of node types to \texttt{add\_serial\_chain} (\ref{code:chain}-\ref{code:add-chain}). To connect the serial chain with the input node, we pass the node indices returned by the node creation methods, \texttt{in\_id} and \texttt{start\_id}, to \texttt{connect} (\ref{code:connect-in}-\ref{code:connect-mix}). This chain corresponds to the first upper left row of the full mixing console shown in Figure \ref{fig:example-console}.
Repeating this procedure multiple times (which is omitted) will complete the graph. 
To compute its output audio, we first convert the graph to tensors (\ref{code:tensor}).
Then, we compute the batched node processing schedule and its indices with \texttt{compute\_render\_data} (\ref{code:render}; corresponds to the line \ref{algo:render:schedule}-\ref{algo:render:calculate_index} of Algorithm \ref{algo:rendering}). With the $5$ stereo source signals of length $2^{17}$ (\ref{code:source}), differentiable audio processors (\ref{code:processors}), and the dictionary of parameters (\ref{code:parameters}), we finally obtain the graph output with \texttt{render\_grafx} (\ref{code:render}; the main loop of Algorithm \ref{algo:rendering}).
This \texttt{output} can be used to calculate a loss and perform the gradient descent to match the target mix.

\begin{figure}[t]
    \centering
    \includegraphics[width=.97\columnwidth]{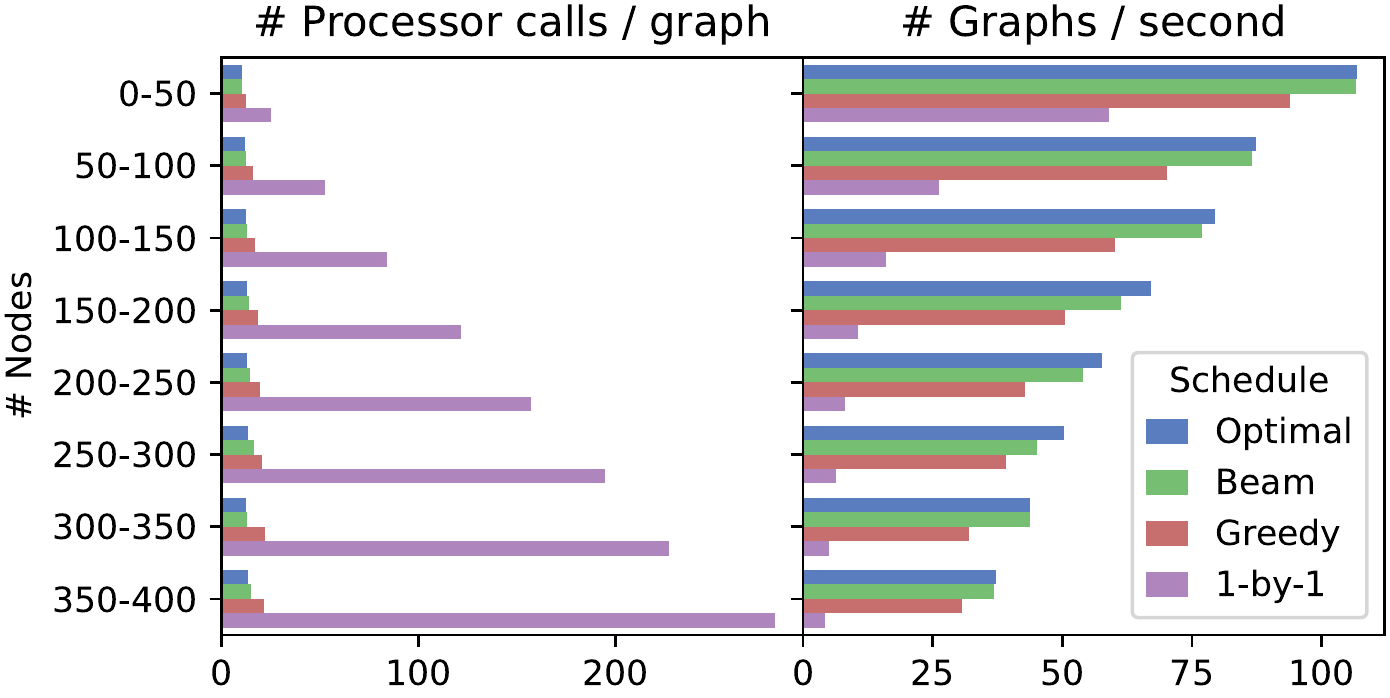}
	\vspace{-1.5mm}
    \caption{\it Speed benchmark results of various scheduling methods.}
    \label{fig:render-speed}
	\vspace{-2.5mm}
\end{figure}

\vspace{1.5mm}
\paragraph{Speed benchmark}
Finally, we evaluated the efficiency of our batched node processing with various scheduling methods.
We benchmarked with a single \texttt{RTX3090} GPU and the \emph{pruned} graphs where negligible processors are removed from the mixing consoles \cite{lee2024searching}.
Figure \ref{fig:render-speed} reports the results.
The optimal schedule achieves $11.8$ processor calls on average. 
The beam ($32$ candidates), greedy, and one-by-one schedules report $12.6$, $16.4$ and $77.6$ calls, respectively.
The one-by-one especially increases the processor calls linearly to the graph size.
Consequently, the batched node processing improves the speed across all graph sizes, especially for large ones. 
While the greedy method performs slightly worse than the optimal, the beam search method closes this gap.

\vspace{-.5mm}

\section{Conclusion}
\vspace{-1mm}
\looseness=-1
Several factors will determine the usefulness of the \texttt{GRAFX} library: usability, flexibility, efficiency of the graph processing algorithms, and diversity of the provided differentiable processors.
Improving these aspects and maintaining the library are left as future work.

\vspace{-1mm}
\vspace{1mm}
\bibliographystyle{IEEEbib}
\bibliography{refs} 
\end{document}